\documentstyle[aps,prl,epsfig,twocolumn,floats,amssymb,amsfonts]{revtex}
\begin{document}
\bibliographystyle{prsty}
\global\firstfigfalse
\global\firsttabfalse
\title{Spin orbit effects in a GaAs quantum dot in a parallel magnetic field}
\author{B.I. Halperin$^{(1)}$, Ady Stern$^{(2)}$,  Y. Oreg$^{(1)}$, J.H. Cremers$^{(1)}$, J.A. Folk$^{(1)}$,  and C.M. Marcus$^{(1)}$}
\address{(1) Lyman Laboratory of Physics, Harvard University, Cambridge,  MA 02138 USA\\(2) Department of Condensed Matter Physics, The Weizmann Institute of Science, Rehovot 76100, Israel
\\
 {\rm \today} \\ \bigskip \parbox{14cm}{ \rm   
We analyze the effects of spin-orbit coupling on fluctuations of the
conductance of a quantum dot fabricated in a GaAs heterostructure. We argue
that spin-orbit effects may become important in the presence of a large
parallel magnetic field $B_{\parallel }$, even if they are negligble for $%
B_{\parallel }=0$. This should be manifest in the level repulsion of a
closed dot, and in reduced conductance fluctuations in dots with a
small number of open channels in each lead, for large $B_{\parallel
  }$. Our picture is consistent with the experimental observations of
Folk et al. \smallskip \\ PACS numbers \vspace{-0.5cm} }}
\maketitle

\smallskip \narrowtext
Recent experiments by Folk et al. \cite{Folk} studied statistics of
fluctuations of the conductance $g$ through a quantum dot in a GaAs
heterostructure with an applied magnetic field $B_{\parallel }$ in the
plane of the sample. In the largest dots studied, the application of $B_{||}$
was observed to reduce the variance of the fluctuations, var($g$), by a
factor of roughly four, in contrast to a reduction factor of two, which was
originaly expected. As noted by Folk et al., the extra reduction might be
understood if, for some reason, spin-orbit coupling increased with the
application of $B_{\parallel }$.

In this note we argue that there should indeed be field-dependent spin-orbit
effects which are unique to a quantum dot, arising from the Rashba and/or
Dresselhaus terms in the effective hamiltonian for electrons in a GaAs
heterostructure. These effects could well explain the observations of Ref. 
\cite{Folk}. Similar effects should appear in the repulsion between energy
levels in a closed dot.

We first explain how spin-orbit coupling and parallel magnetic fields affect
conductance fluctuations. We then show that in quantum dots, unlike open
systems, the effectiveness of spin-orbit interaction grows with increasing $%
B_{||}$. Finally, we examine the consequences of this effect 
on conductance fluctuations in quantum dots, and its relation to the
measurements in Ref. \cite{Folk}.

In a single particle picture, conductance fluctuations through a chaotic or
disordered quantum dot may be crudely understood as arising from
fluctuations in the number of electronic levels in an energy window of size $%
2N\Delta $, and in the matrix elements coupling these levels to the
leads.  Here $\Delta $ is the mean level spacing in the dot, for each
spin state, and $N$ is the number of channels in each lead, (i.e.,
each lead has conductance $2Ne^{2}/h)$.  We assume the leads to be
perfectly coupled to the dot, such that Coulomb blockade effects are
insiginificant. The {\em mean} conductance in this geometry, including
both spin states,
is $\langle g \rangle= Ne^{2}/h$. In the experiments of Ref.
\cite{Folk}, $N$ was in the range 1 to 3.

In the experiments of Ref. \cite{Folk}, a weak perpendicular magnetic field $%
B_{\perp }$ was applied. This field was strong enough to break time-reversal
symmetry for the orbital motion, but not strong enough to produce a
significant Zeeman splitting. Then, if spin-orbit coupling is absent and $%
B_{\parallel }=0$, conductance fluctuations should satisfy 
\begin{equation}
\text{var}(g)=4C_{N},\qquad \qquad ({\mbox{\rm       no spin-orbit, }\quad
B_{||}=0})  \label{fourcn}
\end{equation}
where the constant $C_{N}$ is var($g$) for spinless electrons in a dot with $%
N$ open channels per lead, and the factor $4$ results from the degeneracy of
the two spin states. (From here on we measure all conductances in units of $%
e^{2}/h$, so that $C_{N}$ is dimensionless.) The factor $C_{N}$ depends on
the temperature $T$ through the ratio $T/\Gamma \hbar $ (where $\Gamma
\equiv N\Delta /\pi \hbar $ is the escape rate from the dot) and on the
phase breaking rate $\tau _{\phi }^{-1}$ through the parameter $\Gamma \tau
_{\phi }$. The value of $C_{N}$ can be calculated from random matrix theory,
using the Gaussian Unitary Ensemble (GUE).

For $B_{\parallel }\neq 0$, still in the absence of spin-orbit coupling, the
Fermi levels for spin-up and spin-down electrons are split by the Zeeman
energy $E_{Z}=g^{\ast }\mu _{B}B_{\parallel }$. When $E_{Z}$ is larger than
both $T$ and $\Gamma $, the contributions from the two spin states become
statistically independent, giving
\begin{equation}
\text{var }(g)=2C_{N}\qquad {\mbox{\rm       (no spin-orbit, }}\quad
E_{Z}\gg \Gamma ,T).
\end{equation}

However, in the presence of a strong spin-orbit coupling, the two spin
levels will be mixed, and will be described by a single GUE, with mean
level-spacing $\Delta /2$, and $2N$ open channels in each lead. (Recall that
Kramers degeneracy is already broken by $B_{\perp }$.) Thus in that case, 
\begin{equation}
\text{var }(g)=C_{2N}\qquad \qquad ({\mbox{\rm       strong spin-orbit }}).
\label{ctwon}
\end{equation}

The cross-over to strong spin-orbit coupling should be controlled by
the dimensionless parameter $\lambda =\epsilon _{\rm so}/\Delta $
where $\epsilon _{\rm so}$ is the root-mean-square (RMS) value of the
matrix element $\left\langle i\right| H_{\rm so}\left| j\right\rangle
.$ Here the states $i,j$ have opposite spin directions and orbital
energies that differ by $E_{Z}.$ (The matrix element is to be
calculated with the dot isolated from the leads. The same parameter
$\lambda $ controls the repulsion between levels of opposite spins in
the closed dot.) Then in the presence of $B_{||}$ we can write
\begin{equation}
\text{var }(g)_{B_{||}}=F_{N}(\lambda ,T/\hbar \Gamma ,\Gamma \tau _{\phi }),
\end{equation}
where $F_{N}\rightarrow 2C_{N}$ for $\lambda \rightarrow 0$, and $%
F_{N}\rightarrow C_{2N}$ for $\lambda $ sufficiently large. Note that $%
\Gamma $ is unchanged if $N$ is doubled and $\Delta$ is halved, 
so $\Gamma$ remains constant as one varies $\lambda$. We shall also see that
at least approximately, $C_{N}\approx C_{2N}$, so that $F_{N}$ decreases by
a factor of two as $\lambda $ varies from 0 to $\infty $. Then, if the
system parameters are such that $\lambda $ grows from zero to a large value
as a parallel field $B_{||} $ is turned on, the factor-of-two reduction in $%
F $, combined with the the factor-of-two reduction on breaking the spin
degeneracy, should lead to overall reduction of a factor of $\approx 4$ in
var($g$), relative to the $B_{||} = 0 $ value, in 
Eq. (\ref{fourcn}). This is in accord with the observation of Ref. \cite
{Folk}.

A variety of evidence, based on Random Matrix Theory and other
approaches, suggests that for a large $N$, $C_{N}$ is independent of
$N$, for any fixed value of $\Gamma$, $T$, and $\tau _{\phi }$\cite
{Jalabert1994,Efetov1995,Huibers,Baranger1995,Brouwer1995}.
The biggest deviation from this is presumably for $N=1$ and no
dephasing. At $T=0$, with no dephasing, the value of $C_{N}$ is known,
within Random Matrix Theory, to be $(16-4N^{-2})^{-1}$
\cite{Jalabert1994}. Thus for $N=1$,
the reduction factor $F_N(0,0,\infty)/F_N(\infty,0,\infty)$ is 5/2 
rather than 2.

We define a crossover value $\lambda _{c}$ where var$\left( g\right) $ is
half-way between the values for $\lambda =0$ and $\lambda $ large. We may
estimate $\lambda _{c}$ as the value of $\lambda $ such that $\tau
_{\rm so}^{-1}=\tau _{\phi }^{-1}+\Gamma $, where $\tau _{\rm so}^{-1}=2\pi \lambda
^{2}\Delta /\hbar $ is the rate for spin flip scattering due to the
spin-orbit coupling, given by Fermi's Golden Rule. Writing $N_{{\rm eff}%
}\equiv N+\pi \hbar (\Delta \tau _{\phi })^{-1}$, this gives 
\begin{equation}
\lambda _{c}\approx 0.23N_{{\rm eff}}^{1/2}.  \label{lambdac}
\end{equation}
Numerical calculations, discussed further below and illustrated in Fig. [1], are at least qualitatively
consistent with this estimate, but suggest that the factor 0.23 should be
replaced by 0.1.

Spin-orbit coupling in GaAs heterostructures originates from the asymmetry
of the potential creating the 2DEG (Rashba term) and from the lack of
inversion symmetry in the GaAs lattice structure (Dresselhaus term). The
operator describing the spin-orbit coupling is composed of both terms: 
\begin{equation}
 H_{\rm  so}=\gamma (\vec{v}\times \vec{\sigma})\cdot \hat{z}+\eta (v_{x}\sigma
_{x}-v_{y}\sigma _{y}),  \label{Hso}
\end{equation}
where $\vec{v}$ is the velocity operator, $\vec{\sigma}$ are the Pauli spin
matrices, $\gamma $ and $\eta $ are coupling constants, and we ignore terms $%
\propto v^{3}$. We assume that the 2DEG is grown on a [001] GaAs plane and $x,y$ denote the cubic axes in the plane. In an open clean system, Eq. (\ref{Hso}) leads to a small
spin-splitting of the conduction bands. 
Note that this spin-orbit coupling is different from the one encountered in
measurements of conductance fluctuations in metals. The latter is induced by
impurities, and is characterized by coupling constants that strongly vary
with position.

In the absence of $ H_{\rm so}$, the eigenstates of the electronic Hamiltonian
are products of a spatial part $\left| a \right\rangle $ and a spin
part $\left| \sigma \right\rangle $, where $\sigma =\uparrow (\downarrow )$
denotes spin parallel (antiparallel) to the Zeeman field $\vec{B}_{\parallel
}$. With $H_{\rm so}$, the mean-square value of the dimensionless spin-orbit
coupling $\lambda $ for states with opposite spins at the Fermi energy is 
\begin{eqnarray}
\lambda ^{2}(E_{Z}) &\equiv &\sum_{a b }
\overline{\left| \left(H_{\rm so}\right)_{a \uparrow, b \downarrow} \right|^2} \delta \left( \epsilon
_{a }-\epsilon _{b }-E_{Z}\right) \delta \left( \epsilon _{a
}-\epsilon _{F}\right)
\end{eqnarray}
where $\epsilon _{a },\epsilon _{b }$ are the orbital energies of the
states $\left| a \right\rangle $ and $\left| b \right\rangle $ (i.e.,
the energies at $B_{||}=0$), the over-bar denotes averaging over
disorder, and $\epsilon _{F}$ is the Fermi energy. Here and henceforth
$ O_{s1,s2} \equiv\left\langle s1 \left|  O \right| s2
\right\rangle$. As we now show, in a quantum dot the typical matrix
element $(H_{\rm so})_{a \uparrow, b \downarrow} $ depends on the energy
difference $\epsilon _{a }-\epsilon _{b }$, so that $\lambda ^{2}$
does indeed depend on $E_{Z}.$

For simplicity, we first discuss the case where $\eta =0$ in $H_{\rm so}$, and
we choose $\vec{B}_{\parallel }\parallel \hat{x}$. For a macroscopic system
in the diffusive regime, comparing the Kubo-Greenwood formula with the Drude
formula, one finds 
\begin{equation}
\overline{\left| (v_{x})_{ab} \right| ^{2}}\approx \frac{2D\Delta }{\pi \hbar }\frac{1}{1+(\omega \tau
)^{2}}  \label{diffusive-me}
\end{equation}
where $\hbar \omega \equiv \epsilon _{a }-\epsilon _{b }$, $\ \tau $
is the transport lifetime, and $D=v_{F}^{2}\tau /2$ is the diffusion
constant. Thus, for a large diffusive system and for $\omega \tau \ll
1$, we find$\ \lambda \approx \frac{\gamma v_{F}}{\Delta }\left(
  \frac{\Delta \tau }{\pi \hbar }\right) ^{1/2}$, which does not
depend on $\omega .$ A parallel field does not affect the strength of
spin-orbit coupling as long as $E_{Z}\ll \hbar /\tau $.

In contrast, the confinement of the electron to a quantum dot suppresses the
velocity matrix elements when $\omega \tau _{R}\ll 1,$ where $\tau _{R}$ is 
a Thouless time, which we define as the time for an electron to cross
from the center to the edge of the dot. This is most easily seen if we
use the relation $\left| (v_{x})_{ab}\right| =\omega \left| x_{ab}
\right| $, and note that the matrix element of $x$ is bounded by the
maximum radius $R$ of the dot. More precisely, we may use the relation
\begin{eqnarray}
\overline{\left| x_{ab} \right|^{2}}
&=&\sum_{a }\sum_{b \neq a }\left|x_{ab}\right|^{2}\Delta ^{2}\,\delta \left( \epsilon
_{a }-\epsilon _{b }-\hbar \omega \right) \delta (\epsilon _{a
}-\epsilon _{F})  \nonumber \\
&=&\int_{0}^{\infty }\frac{\Delta \ dt}{\pi \hbar } \left[ \delta x^{2}-\frac{1}{2}\left( x(0)-x(t)\right) ^{2}\right]_{aa} \cos \omega t
\end{eqnarray}
where the last line should be averaged over all states at the Fermi energy; $%
\left(\delta x^{2}\right)_{aa} \approx R^{2}$ is the position uncertainty in the
state $a $; and $\left[ \left( x(0)-x(t)\right) ^{2} \right]_{aa} $
may be approximated by averaging over the appropriate classical
trajectories.

For a dot in the diffusive regime, where $v_{F}\tau \ll R$, we have $\tau
_{R}=R^{2}/2D$. Then, $\left[ \left( x(0)-x(t)\right)
^{2}\right]_{aa} =v_{F}^{2}t^{2}/2$ as long as $t\ll \tau $;
it grows as $2Dt$ for $\tau <t<\tau _{R}$; and finally approaches $%
2\left( \delta x^{2} \right)_{aa}$ for $%
t>\tau _{R}$. Thus, 
\begin{equation}
\overline{\left| \left(v_{x}\right)_{ab}\right|^{2}}\approx \left\{
\begin{array}{lcr}
 c\frac{2D\Delta }{\pi \hbar }(\omega \tau _{R})^{2} &
\text{ for } &
\omega < \tau^{-1}_{R}  \\ \\
c\frac{2D\Delta }{\pi \hbar } &
\text{ for }
& \tau _{R}^{-1}<\omega <\tau ^{-1}
\end{array}
 \right.
  \label{velocity-me}
\end{equation}
where $c$ is a constant which depends on the dot's shape. For a roughly
circular dot of radius $R$, $c\approx O(1)$, and we ignore it below. The
value of $\overline{\left| \left( v_{x}\right)_{ab} \right| ^{2}}$ falls
 off according to (\ref{diffusive-me}) as $%
\omega $ increases further.

For a ballistic chaotic dot, the time scales $\tau $ and $\tau
_{R}\equiv R/v_{F}$ \ coincide, and $D \approx v_{F}R/2$. The second
line of Eq. (\ref {velocity-me}) does not apply. The maximum value of
$\overline{\left| \left( v_{x} \right)_{ab} \right| ^{2}}$ is $\approx
v_{F}R\Delta /\pi \hbar $, obtained when $\omega \tau _{R}\approx 1$.
From these results, we may calculate $\varepsilon _{\rm so}$ and $\lambda
^{2}(E_{Z})=\gamma ^{2}\overline{\left| (v_{x})_{ab} \right| ^{2}}/\Delta
^{2}$.\thinspace\ For
both types of dots the confinement leads to a $B_{||}$ dependence of $%
\lambda ^{2}(\omega )$ for $E_{Z}\tau _{R} \ll \hbar $ (see the inset of Fig.~1). The maximum value of $%
\lambda ^{2}$ is $\lambda _{{\rm max}}^{2} \approx \frac{\gamma ^{2}v_{F}R}{%
\pi \hbar \Delta }.$ 

If $\gamma $ and $\eta $ are both non-zero, $\lambda $ depends on the
direction of $B_{\parallel }$ within the $x-y$ plane. It is different, e.g., for $B_{||}||(110)$ and $B_{ ||}||(1\overline{1}0)$, even for a dot
which is roughly circular. However, the average of $\lambda ^{2}$ over all
directions of $B_{||}$ will be  $(\gamma ^{2}+\eta ^{2})\overline{%
\left|  (v_{x})_{ab} \right|
^{2}}/\Delta ^{2}$.

In order to make a comparison to the experiments of Ref. \cite{Folk},
we consider a ballistic dot where $\gamma ,\eta $ are such that
$\lambda (\Delta )\ll \lambda _{c}$ and $\lambda _{{\rm max}}$ is
greater than the crossover value $\lambda _{c}$ (see Fig. I). At low
temperatures $T\ll \hbar /\pi\tau _{R}$,
the variance var$(g)$ should decrease in two stages as the Zeeman energy $%
E_{Z}$ is increased. In the first it would drop from $4C_{N}^{0}$ to $%
2C_{N}^{0}$ over the range $0<E_{Z}< \pi \max (T,\hbar \Gamma ),$ due to the
removal of spin degeneracy of the levels. Then var($g$) would drop by an
additional factor of approximately $2,$ resulting from the turning on of
spin-orbit coupling, over the larger range $\pi \max (T,\hbar \Gamma
)<E_{Z}<\hbar /\tau _{R}$. (For still larger value of $E_{Z}$, the
conductance fluctuations would increase again.) In Ref. 
\cite{Folk}, $T$ was comparable to $\hbar /\pi \tau _{R}$. Under such
conditions we expect the factor-of-four decrease in var($g$) to occur
smoothly over the range $0<E_{Z}<\hbar /\tau _{R}$. (Recall that
time-reversal invariance is broken by $B_{\perp }$ in all cases).

A quantitative comparison of this scenario to the experiment of Ref. \cite
{Folk} requires information regarding the strength of spin orbit coupling,
which we parametrize by the dimensionless parameter $Q_{so}\equiv (\gamma
^{2}+\eta ^{2})^{1/2}v_{F}/E_{F}$. In terms of $Q_{so}$, for a ballistic dot 
$\lambda _{{\rm max}}\approx Q_{so}N_{e}^{3/4}/3$, where $N_{e}$ is the
number of electrons in the dot. For our scenario to be consistent with the
experiment, we need $\lambda _{{\rm max}}$ to be at least comparable to $%
\lambda _{c}\approx 0.2N_{{\rm eff}}^{1/2}$ for the large measured dot
(where $N_{e}=16,000$ and $N_{{\rm eff}}\approx 6)$ and smaller than $%
\lambda _{c}$ for the small measured dot (where $N_{e}=2,000$ and $N_{{\rm %
eff}}\approx 6).$ These requirements suggest $5\cdot 10^{-3}>Q_{so}>10^{-3}$ .
There are additional uncertainties, however, because our application to 
ballistic chaotic systems of formulas derived for diffusive systems
(e.g., Eq. \ref{velocity-me}) 
involved several unknown numbers of order unity.

Although  $\gamma $ and $\eta $ have been measured previously in other GaAs 
heterostructures, the parameters depend on details of the structure, and are 
difficult to extrapolate from one system to another.  Values of $Q_{so}$ 
extracted from existing data on GaAs 2DEGs include
$Q_{so}\approx 1.6\times 10^{-2}$, from optical measurements\cite{Jusserand},
in a sample with $n=4\times 10^{11}$ cm$^{-2}$, and $Q_{so}\approx 5\times 
10^{-3}$,
from Shubnikov-de Hass measurements\cite{Ramvall}, in a sample with 
$n=1.2\times 10^{12}$ cm$^{-2}$. 
Magneto-resistance measurements \cite{magnetores} in 2DEGs
extract the spin-orbit scattering
rate by studying the crossover from weak localization to weak
anti-localization as the density is increased. At the densities 
where the crossover occurs, typically around $n\sim 6\times 10^{11}$cm$^{-2},$
values around $\tau _{so}^{-1}\sim 4\times 10^{10}$sec$^{-1}$ are found in
the moderate mobility samples ($l\sim 0.5\mu $m), corresponding to $%
Q_{so}\sim 4\times 10^{-3}$.
Our estimates of $Q_{so}$ for the samples of 
Ref. \cite{Folk}, which had $n=2\times 10^{11}$ cm$^{-2}$,
are not incompatible with the range of previous measurements.

The suppression of spin-orbit matrix elements by the confinement to a dot
affects also the scattering rate due to spin-orbit coupling. Note that spin-orbit
scattering processes do not necessaruly result in a spin flip of the electron. The
probability of a spin-flip in a spin-orbit scattering process depends on the
ratio $\gamma/\eta$ and on the intial direction of the spin. We focus on the
case $\eta=0$ and initial spin state in the $x-y$ plane, in which half of
spin-orbit scattering processes involve a spin-flip. We also set $E_Z=0$ for this part of the discussion. The rate of spin-flip due to a spin-orbit scattering process of a state $|a\rangle$ is $\frac{\hbar }{\tau _{\rm so}}=%
\mathop{\rm Im}%
\Sigma (a ,\varepsilon _{a}),$ where $\Sigma (a ,\varepsilon _{a})$
is the on-shell self energy of the state $\left| a \right\rangle $ due
to spin-orbit scattering events, irrespective of the final spin-state. 

To second order in the spin-orbit interaction, the ensemble averaged
imaginary part of the self energy depends on $\left| a \right\rangle $
through $\varepsilon _{a }$. Since the 
total ${\rm Im} \Sigma$ is twice the contribution from the spin flip process, we have
\begin{equation}
\mathop{\rm Im}%
\Sigma (a ,\varepsilon )=2\pi \sum_{b }\overline{\left|
 \left( H_{\rm so} \right)_{a \uparrow, b\downarrow} \right| ^{2}}\delta \left( \epsilon -\epsilon _{b }\right).
\label{self-energy-2-order}
\end{equation}
We neglect the real part of the self energy, which does not significantly
affect our considerations.

Due to the finite escape rate $\Gamma $ from the dot, which is at least
comparable to the level spacing $\Delta $, the $\delta $-functions in (\ref
{self-energy-2-order}) are broadened enough to allow the sum to be replaced
by an integral. Then, in view of Eqs. (\ref{Hso}) and (\ref{velocity-me}),
the {\it on-shell} self energy $\Sigma (a ,\varepsilon _{a })$ 
{\it vanishes}. We go beyond this order, to a self-consistent self-energy,
where 
\begin{equation}
\Sigma (a ,\varepsilon _{a })=2\int \frac{d\varepsilon _{b }}{%
\Delta }\frac{\overline{\left| \left( H_{\rm so} \right)_{a\uparrow b\downarrow}
\right| ^{2}}}{\varepsilon _{a }-\varepsilon _{b
}-\Sigma (b ,\varepsilon _{a })-i\Gamma }
\label{self-energy-self-cons}
\end{equation}
and approximate the solution of (\ref{self-energy-self-cons}) by
substituting (\ref{self-energy-2-order}) in its right hand side. In the diffusive limit we find,
\begin{equation}
\frac{1}{\tau _{\rm so}}\approx \left\{ 
\begin{array}{ccl}
\frac{\tau _{R}}{\tau _{\rm so}^{\infty }}
\left( \frac{1}{4\pi\tau _{\rm so}^{\infty }}
+\Gamma \right)& \text{for}& \frac{\tau
_{R}}{\tau _{\rm so}^{\infty }}\ll 1;\;\Gamma \tau _{R}\ll 1 \\ 
\frac{1}{\tau _{\rm so}^{\infty }} &\text{for}& \frac{\tau _{R}%
}{\tau _{\rm so}^{\infty }} \gg 1
\end{array}
\right.   \label{so-rate-limits}
\end{equation}
where $\frac{1}{\tau _{\rm so}^{\infty }}=4 \gamma ^{2}D/\hbar^2$ is
the spin-flip scattering rate in an open system.  As expected, when
$\tau _{R}\gg\tau _{\rm so}^{\infty }$ the confinement of electrons to the
dot does not significantly affect spin-orbit scattering rate. In
contrast, for a small dot spin-orbit scattering rate is suppressed.
When $\Gamma \ll \frac{1}{\tau
_{\rm so}^{\infty }}$ it becomes of fourth order in the coupling constants, $%
\frac{1}{\tau _{\rm so}}\approx \frac{\tau _{R}}{4\pi(\tau _{\rm so}^{\infty })^{2}}.$
At this order, the smallness of spin-orbit matrix elements at close energies
is overcome by virtual transitions of high energy difference ($\sim \tau
_{R}^{-1}$).

The suppression of $\tau _{\rm so}^{-1}$ in small dots has implications for
electronic transport through the dots. Normally, for an open system, when $%
\tau _{\rm so}^{-1}$ gets larger than $\tau _{\phi }^{-1}$, weak localization
turns into weak anti-localization, and the magneto-resistance becomes
positive. The corresponding criterion for a quantum dot compares the reduced 
$\tau _{\rm so}^{-1}$ with $(\tau _{\phi }^{-1}+\Gamma )$. In fact, the
spin-orbit scattering rate relevant for transport may be even smaller than
the already reduced $\tau _{\rm so}^{-1}$ given by (\ref{so-rate-limits}). Even
in open 2D electron structures in GaAs, there are subtleties due to the fact
that spin rotations induced by $H_{\rm so}$ are correlated with spatial
displacements of the electron. As a result the spin-orbit relaxation rate
relevant for weak antilocalization and for conductance fluctuations at $%
E_{Z}=0$ can be smaller than that obtained from the above analyses. In fact,
when $\eta =\pm \gamma $, there is actually no weak antilocalization, if the
cubic term, $\propto v^{3}$ in $H_{\rm so}$, is ignored \cite{knap1996}. 

\begin{figure}[h]
  \centerline{\epsfxsize=8.0cm \epsfbox{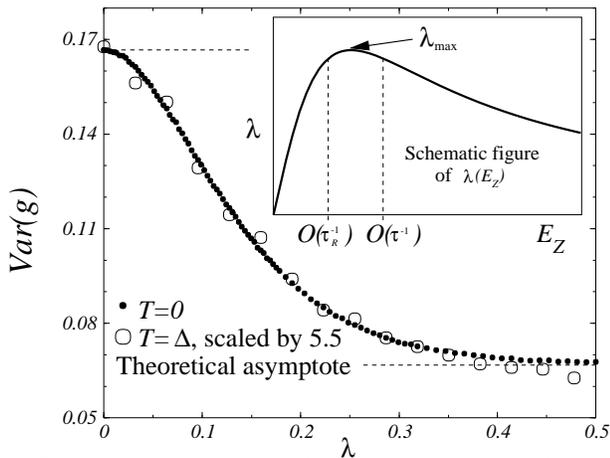}}
\caption{ Variance of conductance for the case 
$N=1,\ \tau_\phi^{-1}=0$, as a function of the
  coupling $\lambda$, for $T=0$ and $T=\Delta $. Asymptotes show known results
for $\lambda=0$ and $\lambda= \infty$ at $T=0$.
Data for $T=\Delta $ were scaled by a factor of
  $5.5$. The inset shows {\em schematic} behavior of $\lambda$ as
  a function of the Zeeman energy $E_Z$, following
 Eqs.~(\protect{\ref{diffusive-me}}) and
    (\protect{\ref{velocity-me}}). }

\end{figure}

Before concluding, we explain the random matrix calculations leading
to Fig.  [1].  These calculations are aimed at studying the $\lambda
$--dependence of var($g $) in the presence of a strong $B_{||}$. The
Hamiltonian of the closed dot was modeled by a $2M\times 2M$ random
matrix $H$ of the form $H_{ij}=\eta _{ij}K_{ij}$ where $1<i<M$ labels
states with spin up, $M+1<i<2M$ labels states with spin down, $\eta
_{ij}=1$ if $i$ and $j$ are states with the same spin, $\eta
_{ij}=\varepsilon _{\rm so}=\pi \lambda /M^{1/2}$ if $i$ and $j$ are
states with opposite spin, and $K$ is a random Hermitian matrix from
the GUE, with the distribution
$
P(K)\propto e^{-\frac{1}{2}\,{\rm tr}\,K^{\dagger }K}.
$
This matches the definitions in the text because $\varepsilon _{\rm
  so}$ is the RMS value of the matrix element connecting two states in
the different spin blocks, and $\Delta =\pi /M^{1/2}$ is the average
level spacing at the center of the band for one block. For the case
$N=1$, we connect the system to ``leads'' with perfect conducting
channels at 2 sites for spin up and 2 sites for spin down, and
calculate the $2\times 2$ transmission matrix $t$ for energies near
the center of the band\cite{Verbaarschot}.

At $T=0$ one finds the conductance by using the Landauer formula $g=
{\rm tr}\,tt^{\dagger }.$ To obtain results at finite $T$, for
each realization of the random matrix we first calculate the transmission
matrix and thus the $T=0$ conductivity $g^{(0)}(E)$ for a
range of energies $E$. This conductance is weighted by the derivative of the
Fermi function and integrated to give $g(T)=\int_{-\infty }^{\infty }\frac{df%
}{dE}g^{(0)}(E)dE.$

The conductance fluctuations as a function of $\lambda$ for zero
temperature and temperature $\Delta$ are shown in Figure 1. The data for $%
T=\Delta$ were accumulated from 5000 realizations with $M=60$, while the
data for $T=0$ were obtained from $10^6$ realizations with $M=20$. More
limited calculations at $T=0$ with $M=60$ showed differences of less than 10
\% from $M=20$.

It is clear from the figure that the $T=\Delta $ results have the same
depedence on $\lambda $ as the $T=0$ results, and we have evidence that this
remains true for higher T. The reduction in var($g$) by a factor $\approx 6$
is consistent the theoretical expectation\cite{Huibers} that $C_N(T) \approx
C_N(T=0) \pi \hbar \Gamma / 6 T$, for $T \geq \pi \hbar \Gamma$, if
dephasing is absent. The middle of the crossover occurs roughly at $\lambda
=0.1$, which is somewhat less than the value 0.23 given by Eq. (\ref{lambdac}%
).

Calculations for $2\leq N\leq 4$ at $T=0$ (not shown) are consistent with a
crossover value $\lambda _{c}$ scaling as $N^{1/2}$, as predicted by (\ref
{lambdac}). Dephasing can also be included by using a ``third-lead'' model
of the type discussed by Brouwer and Beenakker\cite{Brouwer1997}. Results
for $N=1$ and $T=0$ seem to show a variation of $\lambda _{c}$ somewhat
slower than $N_{{\rm eff}}^{1/2}$, at least in the range $1<N_{{\rm eff}}<4$.

In summary, we presented a theory by which the effect of spin-orbit on
conductance fluctuations in a quantum dot depends strongly on an applied
parallel magnetic field. This theory may well explain the experimental
observation of Ref. \cite{Folk}.

We gratefully acknowledge discussions with P.W. Brouwer and I. Aleiner, and
financial support from NSF grants DMR99-81283 and DMR97-14725, ARO grants 341-6091-1-MOD 1 and DAAD19-99-1-0252, the US-Israel BSF,
DIP-BMBF, the Israel Academy of Science, the V. Ehrlich chair and the DoD.


\vspace{-0.5cm}
\begin{references}
\vspace{-1.5cm}
\bibitem{Folk}  J. A. Folk, S. R. Patel, K. M. Birnbaum, C. M. Marcus, C. I. Duruoz 
and J. S. Harris, Jr., Phys. Rev. Lett. (in press) (cond-mat/0005066).

\bibitem{Jalabert1994}  R. A. Jalabert, J.-L. Pichard and C. W. J.
Beenakker, Europhysics Letters{\bf ~27}, 255 (1994).
H.~U. Baranger and P.~A. Mello, Phys. Rev. Lett. 
{\bf 73}, 142 (1994).

\bibitem{Efetov1995}  K.B. Efetov, Phys. Rev. Lett.{\bf 74}, 2299 (1995).

\bibitem{Huibers}  A.G. Huibers, S.R. Patel, C.M. Marcus, P.W. Brouwer, C.I.
Duru\"{o}z and J.S. Harris, Jr., Phys. Rev. Lett. {\bf 81}, 1917 (1998).

\bibitem{Baranger1995}  H.~U. Baranger and P.~A. Mello, Phys. Rev. B {\bf 51}%
, 4703 (1995).

\bibitem{Brouwer1995}  P. W. Brouwer and C. W. J. Beenakker, Phys. Rev. B%
{\bf ~51}, 7739 (1995).

\bibitem{Brouwer1997}  P.~W. Brouwer and C.~W.~J. Beenakker, Phys. Rev. B 
{\bf 55}, 4695 (1997).

\bibitem{Jusserand}  B.~Jusserand, D.~Richards, G.~Allan, C.~Priester, and B.~Etienne, 
Phys. Rev. B{\bf 51}, 4707 (1995).

\bibitem{Ramvall}  P.~Ramvall, B.~Kowalski, and P.~Omling, Phys. Rev. B{\bf %
55}, 7160 (1997).

\bibitem{magnetores}  O. Millo et al, Phys. Rev. Lett. {\bf 65} 1494 (1990),
J. E. Hansen, R. Taboryski, and P. E. Lindelof Phys. Rev. B {\bf 47}, 16040 (1993), P. Dresselhaus et al,
Phys. Rev. Lett. {\bf 68} 106 (1992).

\bibitem{Verbaarschot}  J.J.M. Verbaarschot, H.A. Weidenm\"{u}ller and M.R.
Zirnbauer, Phys. Rep. {\bf 129}, 367 (1985).

\bibitem{knap1996}  W. Knap et al., Phys. Rev. B {\bf 53}, 3912 (1996).
\end{references}
\end{document}